# Hysteresis and mode switching in FIB engineered random laser diodes

A. Martínez Pacheco[1], S. Rodríguez Corvillo[2], A. Rivera-Calzada[2], A. Consoli[1,3], J. Santamaría[2] and C. López[1]

*[1]Instituto de Ciencia de Materiales de Madrid (ICMM-CSIC), Sor Juana Inés de la Cruz 3, 28049 Madrid, Spain*
*[2]GFMC, Departamento de Física de Materiales, Universidad Complutense de Madrid, Plaza Ciencias, 1, 28040 Madrid, Spain*
*[3]Escuela de Ingeniería de Fuenlabrada, Universidad Rey Juan Carlos, Camino del Molino 5, 28942 Fuenlabrada, Madrid, Spain*
**Corresponding author: antonio.consoli@urjc.es*

**Complex dynamical behavior in laser diodes is commonly induced by external perturbations such as optical or optoelectronic feedback and optical injection. The structural transformation of a stable laser diode into a stand-alone nonlinear dynamical system would provide a practical alternative as a single and compact device. Here we demonstrate the fabrication of such a device by milling, via focused ion beam, the output facet of a Fabry-Perot laser diode with a predesigned mask, obtaining a scattering surface which provide feedback for random lasing. The fabricated random laser diode shows multi-mode emission with current dependent spectral reconfigurations, hysteresis in output power and spectrum and mode switching at constant bias current. We attribute this behavior to the interplay between scattered optical feedback and carrier induced local variations of refractive index and temperature at the output facet surface, where defects are located. The observed phenomena reveal complex nonlinear behavior emerging from a single standalone device, which could be tailored by varying the milling pattern, thereby providing a flexible platform for different applications.**

Semiconductor lasers, while stable when operated in isolation, can exhibit complex dynamical behavior, including self-pulsation, bistability, and chaos, when subjected to external perturbations such as optical or optoelectronic feedback and optical injection[1,2]. Complex dynamics can also arise intrinsically in free-running devices, including large-area VCSELs[3], broad-area lasers[4,5], ring resonator[6] and chaotic cavities[7], where spatial effects, transverse-mode interactions, and nonlinear mode competition can destabilize steady-state emission.

Complex dynamical behavior in laser diodes (LDs) has found applications in areas ranging from secure communications[8], random number generation[9] and chaotic lidar[10] to microwave photonics[11] and sensing[12]. Recently, such systems have emerged as promising hardware platforms for neuromorphic photonics[13], optical information processing[14] and reservoir computing[15,16], where nonlinear transformation and fading memory are key computational resources.

The common underlying mechanism in these devices and configurations is the nonlinear coupling between the optical field, the carrier population, and the refractive index, which can destabilize steady-state emission and give rise to complex dynamical regimes.

Random lasers (RLs)[17,18], in which optical feedback arises from scattering elements, show the potential for an alternative route to complex laser dynamics, owing to their typical large number of irregular spatial and frequency modes and competition phenomena that can arise[19,20]. Chaotic behavior has been observed in optically pumped RLs based on dye-doped scattering media[21], although these systems rely on optical pulsed excitation.

Although several electrically pumped random laser diodes (RLDs) with spatially distributed optical feedback have been demonstrated[22–27], these studies have primarily focused on device fabrication and steady-state emission characteristics, such as lasing threshold, output power, and spectral properties, for random lasing demonstration[28]. This emphasis likely reflects the considerable challenge of fabricating and demonstrating such devices, which typically requires introducing a random refractive-index distribution within the active region during the fabrication process. As a result, their nonlinear dynamical behavior has received comparatively little attention.

An alternative approach is to introduce scattering feedback at the boundaries of the active region, giving rise to RLs with spatially localized, or non-distributed, feedback[29–34], avoiding the need to engineer disorder within the active region.

RLDs with non-distributed feedback have been obtained by introducing random scattering defects onto the output mirror of a Fabry–Pérot (FP) LD, via pulsed laser ablation[35,36]. Although this approach simplifies device fabrication, the uncontrolled size and spatial distribution of the resulting defects limit reproducibility.

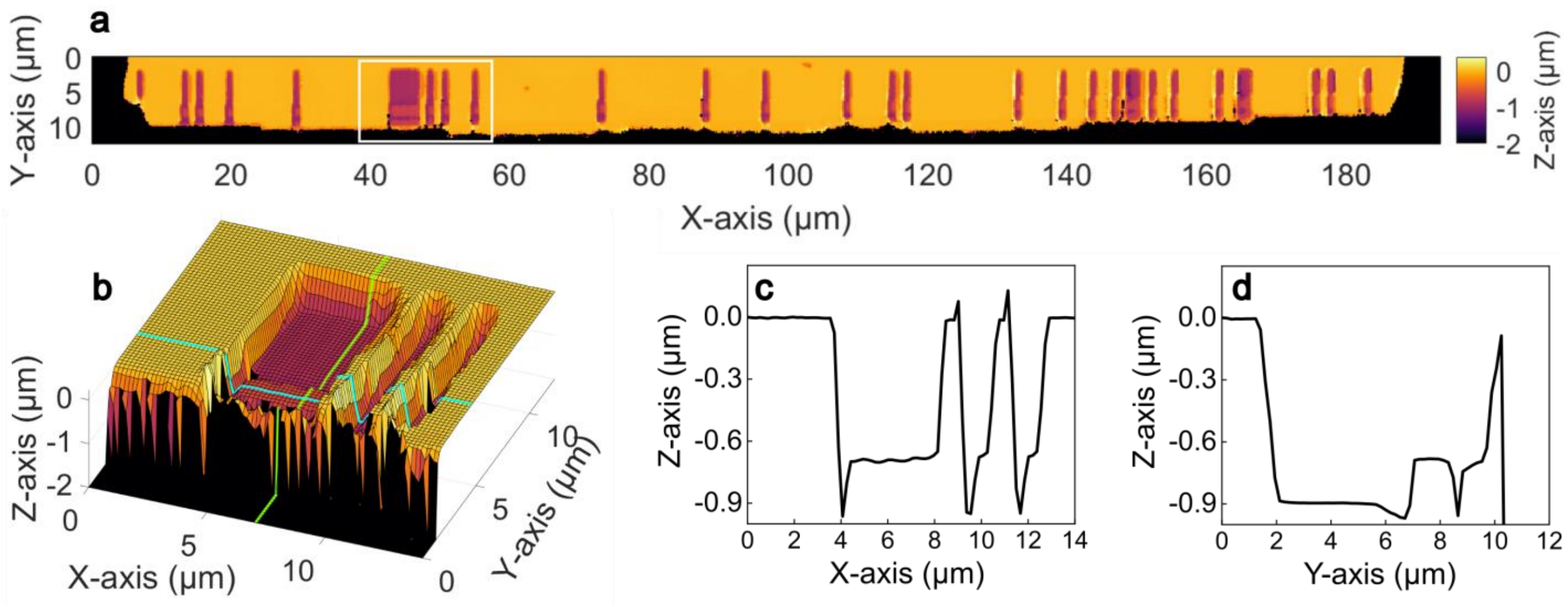

**Figure 1. Output mirror surface characterization.** 2D surface map of the FIB-milled output facet along the entire 180 µm width (a). 3D surface topography of a representative milled slot region, highlighted with white rectangle in Fig 1a (b). Cross-sectional depth profile extracted along the X-axis (c) and the Y-axis (d), corresponding to cyan and green lines in Fig. 1b, respectively.

Here, we fabricate and study RLDs obtained by focused ion beam (FIB) milling of the output facet of a FP LD, according to a predefined mask. The resulting devices exhibit hysteresis in optical power and spectrum, with stochastic variations in transition currents, current-dependent spectral reconfigurations, and unstable pulsing regimes characterized by modal competition. These observations suggest that the device acts as a complex nonlinear dynamical system in which disordered optical feedback intrinsically couples the optical field to the carrier and thermal dynamics.

Remarkably, this rich phenomenology emerges in a single, monolithic device, without external cavities, optical injection, or multisection architectures. Dynamical features such as bistability, long-lived emission configurations and stochastic mode switching point to the potential for accessing multiple interacting states and history-dependent responses through a simple, disorder-engineered structure. Beyond their fundamental interest, these properties may provide opportunities for exploring compact photonic platforms for neuromorphic computing and reservoir computing, where nonlinear state transitions and memory effects are key ingredients.

## Device fabrication

Commercially available AlGaInP LDs (Thorlabs HL6388MG) were used as the original devices, a full description of the original LDs is given in Section 1 of Supplementary Information. These LDs emit at 635 nm, exhibiting a typical threshold current $I_{TH}$ = 90 mA and efficiency slope $\eta_{SLOPE}$ = 0.9 W/A, corresponding to a differential quantum efficiency $\eta_D$ = 51 %. Devices are encapsulated in a TO-56 can package which is opened for inspection and device modification. Structurally, the devices feature a transverse multimode FP cavity with dimensions of 1.45 mm (length) along the z-axis and 180 µm (width) along the x-axis, while the die thickness in the epitaxial growth direction (y-axis) is about 100 µm. After chip flip, the AlGaInP multi-quantum-well active region, opposite the growth substrate, is positioned close to the bottom contact at the lower part of the device.

The milling procedure is carried out as follows. A fundamental slot element is first defined with dimensions of 1 µm along the x-axis and 10 µm along the y-axis. The lateral size is selected to be comparable to the emission wavelength, being approximately 1.5 times larger, whereas the vertical dimension is intentionally chosen to largely exceed the thickness of both the multi-quantum-well active region and the optical confinement layers. This ensures effective milling of the emitting region even under possible alignment inaccuracies during the FIB milling process.

A milling pattern is then generated, comprising 180 adjacent slots distributed along the lateral direction in order to cover the entire 180 µm-wide output facet. From these 180 predefined slots, only a fraction is selected for milling according to the desired milling ratio, defined as the ratio between milled and total number of regions. Here, we consider a device in which 32 random slots of the 180 available have been milled, corresponding to a milling ratio of 18%.

The etched patterns were fabricated using a state-of-the-art dual FIB-SEM VELION S system by RAITH. FIB

milling of the slots was carried out with $Au^+$ ions at an acceleration voltage of 35 kV, a 40 μm aperture, and a beam current of 290 pA. The milling depth is adjusted by controlling the ion dose delivered by the FIB system. To establish the correlation between ion dose and etched depth, a sacrificial device is first employed for calibration. Based on this calibration, the ion dose (0.8 nC/μm$^2$) is chosen to produce a milling depth of approximately 1 μm.

Although different milling ratios and slot depths can be implemented, we restrict the present study to a single milling ratio and a fixed slot depth for simplicity.

## Surface characterization

In Fig. 1a, we show the milled mirror facet, i.e. the emitting area of the active layer located at the bottom of the laser chip. This has been measured with an optical profilometer: the color bar in Fig.1a represents the milling depth along the z-axis. The distribution of the unit slots corresponds to the designed mask, with random distribution of 32 slots. Large, ablated regions correspond to 1 μm-wide milling slots which have, by chance, been placed at adjacent positions. A rectangular region with dimensions of 15 μm (x-axis) and 10 μm (y-axis) is highlighted with white line in Fig. 1a and blown up as a 3D plot in Fig. 1b. The 3D graph in Fig. 1b shows that the ablated areas have defined borders, however the edges and the bottoms of the ablated areas have irregular profiles, with non-uniform material removal on the bottom and material re-deposition on the edges, resulting in rough bottoms and edges of the milled areas.

This is better appreciated in Fig. 1c, where we plot the cross section of Fig. 1b, along the x-axis (cyan line in Fig. 1b). In Fig. 1c, the depth of the milled regions is ~ 0.65 μm. At the edges of the milled regions, we observe abrupt fluctuations of the depth profile, attributed to uneven removal (pits which go below the bottom of the ablated slots) and material deposits (peaks above the mirror surface).

In Fig. 1d, we plot the cross section of Fig. 1b along the y-axis (green line in Fig. 1b), corresponding to an etched region. We observe non-uniform milling in which a small section, between the y-axis positions of 7 μm and 8.3 μm, have a reduced milling depth. This is associated with the active and confinement layers, which having a different material composition than the bulk of the dice, presents a different milling rate.

The profiles of Fig. 1c and 1d are plotted as representative for the entire milled area, with the aim of showing their sub-micrometer irregular features, due to different material etching depths, material re-deposition and imperfections of the FIB milling process.

## CW emission characteristics.

The RLD is placed on a laser diode mount and temperature and current controlled via PC. The output power is measured with a calibrated photodiode and the emission spectrum with a spectrometer with a resolution of 0.15 nm. More details on the experimental set-up are given in Section 1 of Supplementary Information.

In Fig. 2a, we plot the output power and full width at half maximum (FWHM) as a function of current. We observe sudden increase of output power and spectral narrowing, clear clues of threshold crossing in lasing action.

The RLD has a threshold current $I_{TH}$ = 118 mA and an efficiency slope of 0.14 W/A, which corresponds to a differential quantum efficiency, $\eta_D$ = 7 %. The laser linewidth decreases abruptly from 12 nm well below threshold to 0.15 nm just above $I_{TH}$. Spectral width varies above $I_{TH}$ between 0.15 nm and 1.9 nm.

The original LD (see Section 1 in Supplementary Information) has a $I_{TH}$ = 90 mA, efficiency slope of 0.98 W/A and $\eta_D$ = 51 %. The increase in threshold current and the decrease in quantum differential efficiency, $\eta_D$, is attributed to augmented mirror losses, $\alpha_M$, internal scattering losses, $\alpha_I$, and decreased internal quantum efficiency, $\eta_I$. Considering that $\eta_D = \eta_I \cdot \alpha_M/(\alpha_M + \alpha_I)$, the sole reduction in output mirror reflectivity, due to the FIB milling process, would actually increase the output light, thus the slope and quantum differential efficiencies. We attribute the low value of $\eta_D$ in the RLD to augmented internal losses due to backscattered light into the active region outside the guided modes and reduced internal quantum efficiency, due to non-radiative recombination occurring at the mirror surface associated with imperfections, dangling bonds, and ion implantation[37,38].

At 205 mA, we observe a sudden power increase in output power in Fig. 2a, that will be commented on later in the text.

In Fig. 2b, we plot the (normalized) emitted spectrum as a function of bias current above threshold. The spectra are characterized by multiple narrow peaks, with most of the emission intensity concentrated in one or a few of them. The spectral signatures vary abruptly with current, with the dominant modes switching to different spectral positions. Examples of this behavior are shown in Fig. 2c for spectra measured at 166 mA, 204 mA, and 205 mA. The spectra are normalized to their maximum intensity to facilitate comparison of their profiles. To estimate the FWHM linewidth of the full emission spectrum, we perform a numerical fit to a Gaussian profile, thereby minimizing the effect of the spiky spectral structure on the linewidth estimation.

At 166 mA, the spectrum consists of a broad emission background (FWHM = 2.5 nm) with several superimposed narrow peaks, whose spacing is non-uniform and averages approximately at 0.3 nm. At 205 mA, two intense, resolution-limited peaks emerge at 635.5 and 636.6 nm, with linewidths of approximately 0.13 nm, while the broad underlying emission retains a FWHM of 2.5 nm. At 206 mA, the emission is dominated by a single resolution-limited peak at 636.5 nm, superimposed on a smooth background with a reduced FWHM of 1.8 nm.

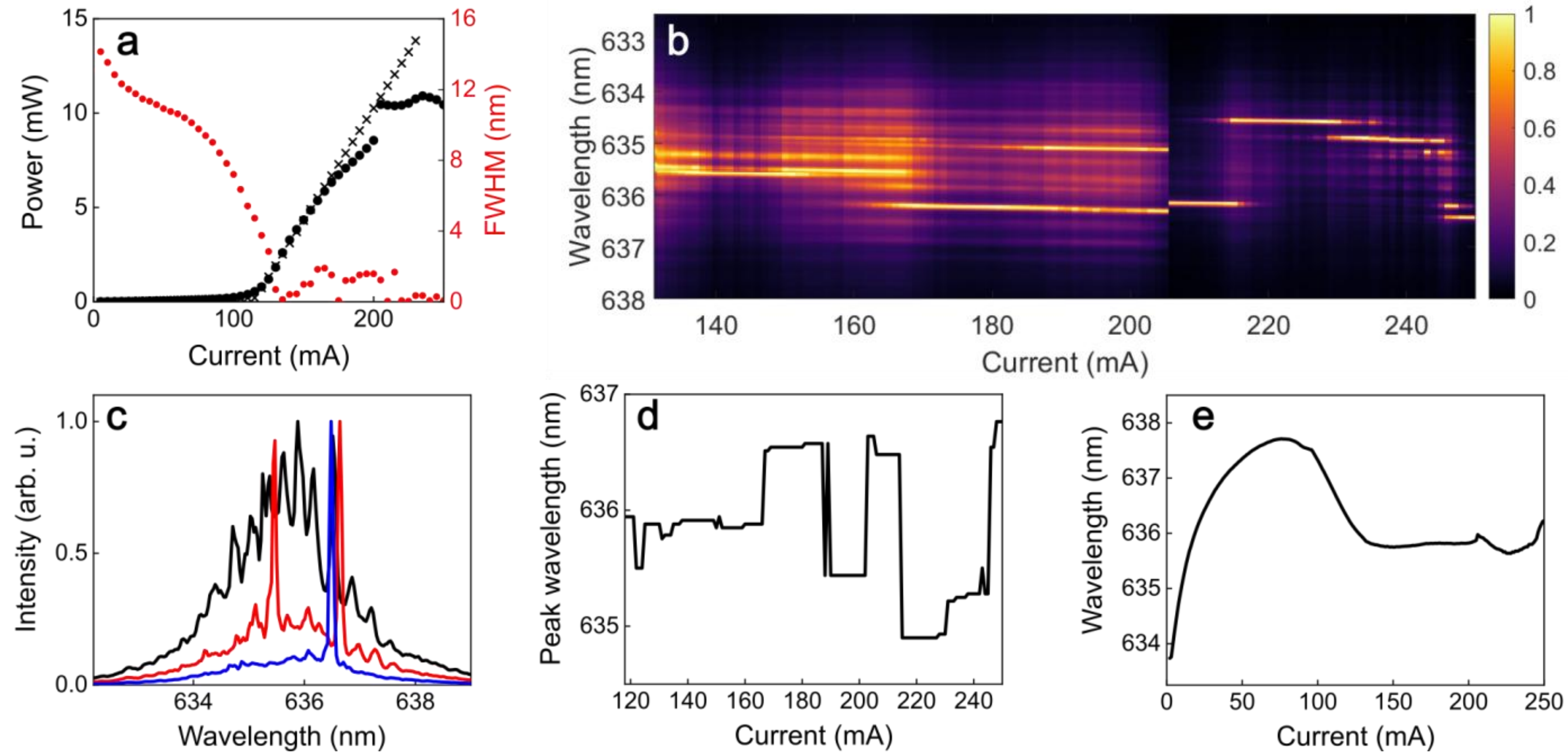

**Figure 2. Continuous-wave output characterization.** Measured (left axis, black dots) and fitted (left axis, black crosses) output power and full-width at half-maximum spectral linewidth (right axis, red dots) as a function of bias current (a). Normalized emission spectra versus bias current, above threshold (from 130 mA to 250 mA) (b). Emission spectra at bias currents of : 166 mA (black line), 205 mA (red line) and 206 mA (blue line) (c). Peak emission wavelength as a function of bias current above threshold (d). Spectral centroid versus bias current across the full current range (e).

This behavior is characteristic of multimode RLs, where the degree of mutual coupling between the modes can vary, giving rise to different emission regimes, usually dubbed as resonant and non-resonant emission regimes. Multiple modes compete for the available gain, resulting in spectra with several peaks and a broad emission background. Depending on the bias current, several modes can coexist with comparable intensity, as observed at 166 mA, while only a few modes dominate at 204 mA, or a single mode dominates at 205 mA.

The measured average peak spacing of ~0.3 nm is significantly larger than the FP free spectral range or longitudinal-mode spacing expected for a cavity length $L$ = 1.45 mm. Using $\Delta\lambda_{FP} = \lambda^2/(2\cdot n\cdot L)$, with central wavelength $\lambda$ = 635 nm and group index $n$=3.5, gives $\Delta\lambda_{FP}$ = 0.04 nm, nearly one order of magnitude smaller than the observed value. The ~0.3 nm spacing is therefore attributed to the strong mode coupling and competition underlying the multimode emission.

In Fig. 2d, we show the wavelength of the strongest peak as a function of the bias current. We observe a predominantly step-like evolution, with the emission wavelength remaining nearly constant over certain current ranges and undergoing abrupt transitions to different spectral positions. These mode hops occur over the full range of bias currents and lead to several distinct quasi-stable emission wavelengths across the investigated range. The absence of a continuous, smooth wavelength shift (customary in ordinary LDs), together with the abrupt switching between modes, is consistent with the strong mode competition observed in the spectra of Fig. 2b and 2c.

The spectral centroid, defined as the intensity-weighted average wavelength of the measured spectrum, is plotted in Fig. 2e as a function of injection current. Below threshold, the centroid progressively redshifts, reaching approximately 637.7 nm at 80 mA, followed by an abrupt blueshift of about 1.5 nm as the threshold region is approached. Above threshold, the centroid remains nearly constant over a broad current range, before undergoing a second blueshift around 220 mA and subsequently increasing up to 250 mA.

This behavior differs markedly from that of a conventional FP LDs, for which the wavelength typically exhibits a progressive redshift below threshold, a small blueshift around threshold, and an approximately linear redshift above threshold (see the original LD characterization in Section 1 of Supplementary

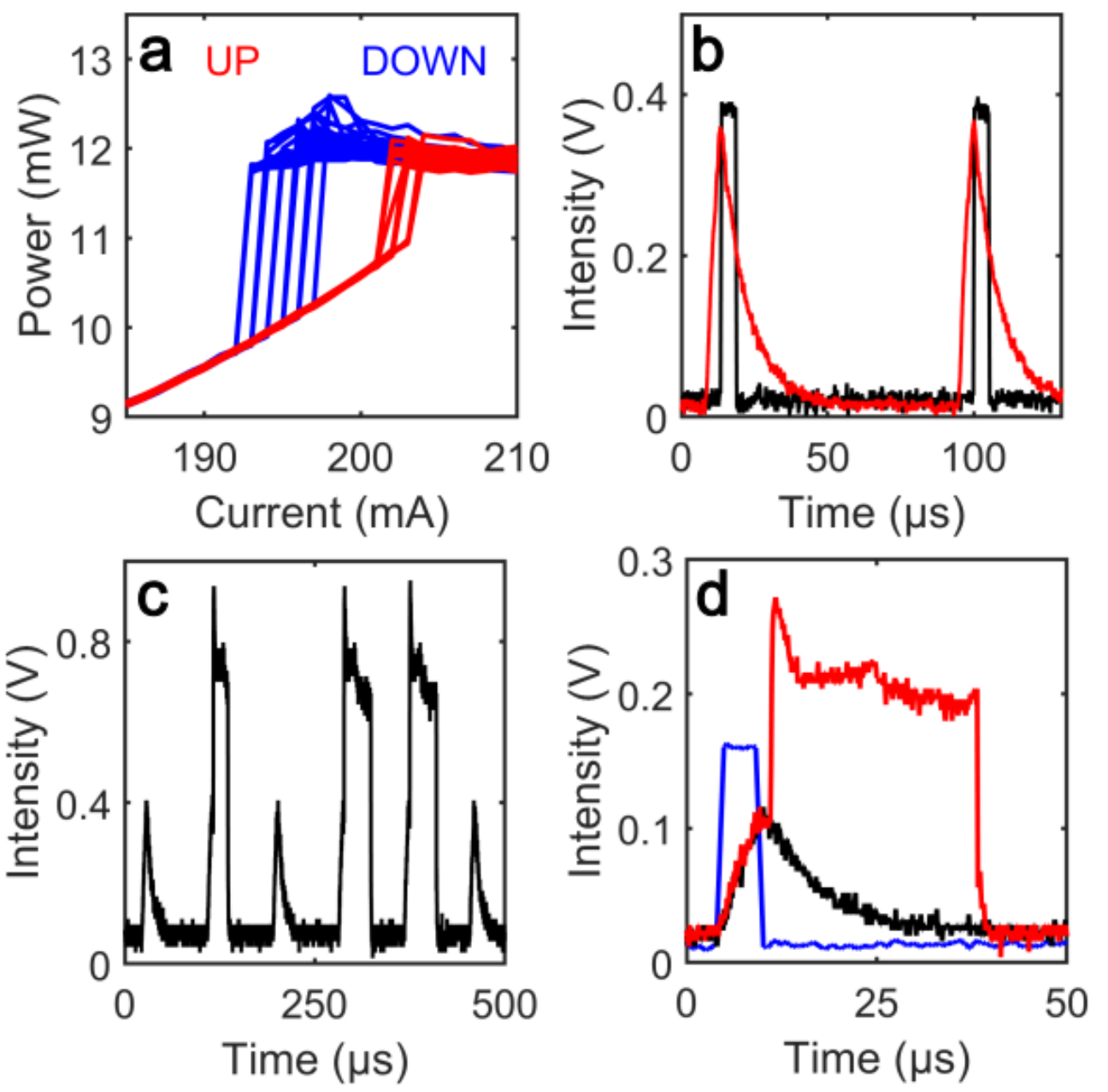

**Figure 3. Hysteresis, pulse excitation, and temporal memory dynamics.** Light output power versus drive current during repeated forward (red) and backward (blue) 100 current sweeps (a). Electrical input (black line) and optical output (red line) for $P_{AMP}$ = 0.4 V (b). Optical output (black line) for $P_{AMP}$ = 0.7V (c). Detail of single-pulse temporal dynamics comparing the electrical input pulse (blue), low-state optical response (black), and high-state optical persistence (red) (d).

Information). The observed deviations indicate a more complex interplay between thermal effects and carrier-induced refractive-index changes in the modified device. Below threshold, the progressive redshift is primarily attributed to junction heating, as common in FP LDs. The pronounced blueshift close to threshold suggests a sudden carrier-induced modification of the refractive index associated with the onset of stimulated emission. Above threshold, the nearly constant centroid can be interpreted as a partial compensation between the thermally induced redshift and the carrier-induced refractive-index contribution.

The blue and red shifts, occurring between 220 mA and 250 mA, coincide with the abrupt spectral transition observed in Fig. 2d and is likewise attributed to a sudden carrier-induced refractive-index change accompanying the transition between different multimode emission regimes.

## Stochastic hysteresis and periodic excitation

The sudden power increase at 205 mA, observed in Fig. 2a, suggests that the device was crossing some kind of threshold. Consequently, we performed several continuous measurements by sweeping up and down the bias current and measuring the emitted power and spectra. In Fig. 3a, we show the output power detected when sweeping upward (red traces) and downward (blue traces), between 185 mA and 215 mA, for 100 measurements.

Across all measured cycles, a well-defined counterclockwise hysteresis loop is observed, indicating the coexistence of two distinct states. As shown in Fig. 3a, the loops feature abrupt power surges (in the upward sweep) and cuts (during the downward current sweep).

Although the transition currents fluctuate randomly from cycle to cycle without any systematic drift, their ensembles, denoted as $I_{DOWN}$ and $I_{UP}$, yield mean values of 200.6 mA and 192.6 mA. This gives an average hysteresis width of approximately 8 mA. Notably, the corresponding standard deviations of 0.6 mA and 1.2 mA reveal an asymmetric width between the up and down transition distributions. The apparent discretization of the switching currents arises from the finite current-step resolution used in the measurements (1 mA) and does not necessarily reflect intrinsic quantization. Emitted spectra undergoes an abrupt change from multimode to a single dominant mode (as seen in Fig. 2, when increasing the current from 205 mA to 206 mA) when the current is increased above $I_{UP}$. When the current is swept back down, the single dominant mode persists until $I_{DOWN}$ is crossed, whereupon multimode emission is restored. See Section 2 of the Supplementary Information for further details.

We investigate the dynamic behavior of the device in the hysteresis region by modulating the RLD with a squared pulse periodic waveform, with pulse duration of 5 μs and period 85 μs, superposed to a DC bias current, $I_{DC}$ = 190 mA, via the current driver. The RLD output is sent to an amplified photodiode and recorded on a digital oscilloscope, see Section 2 of Supplementary Information.

Initially, the pulse amplitude, $P_{AMP}$, is set to $P_{AMP}$ = 0.4 V. The detected signal at oscilloscope corresponds to a train of identical optical pulses with intensity profile characterized by exponential rising and falling edges, see Fig. 3b. This pulse profile is due to the intrinsic maximum modulation bandwidth of the original LD, which acts as a low pass filter for the fast rising and falling edges of the electrical square pulse. With this value of $P_{AMP}$, the RLD is directly modulated between two currents on the upward trace, well inside the hysteresis cycle but not reaching currents where the output power surge occurs ($I_{DOWN} < I < I_{UP}$).

The pulse voltage amplitude is then increased to $P_{AMP}$ = 0.7V, while keeping constant the pulse duration, modulation period and $I_{DC}$ value. In this condition, the current reaches the region where the output power surges ($I_{UP} < I$ ) and we observe the appearance of output pulses of much larger intensity, occurring at random periods of the modulating signal, which alternate with low pass filtered, lower intensity pulses observed with $P_{AMP}$ = 0.4V, see Fig.

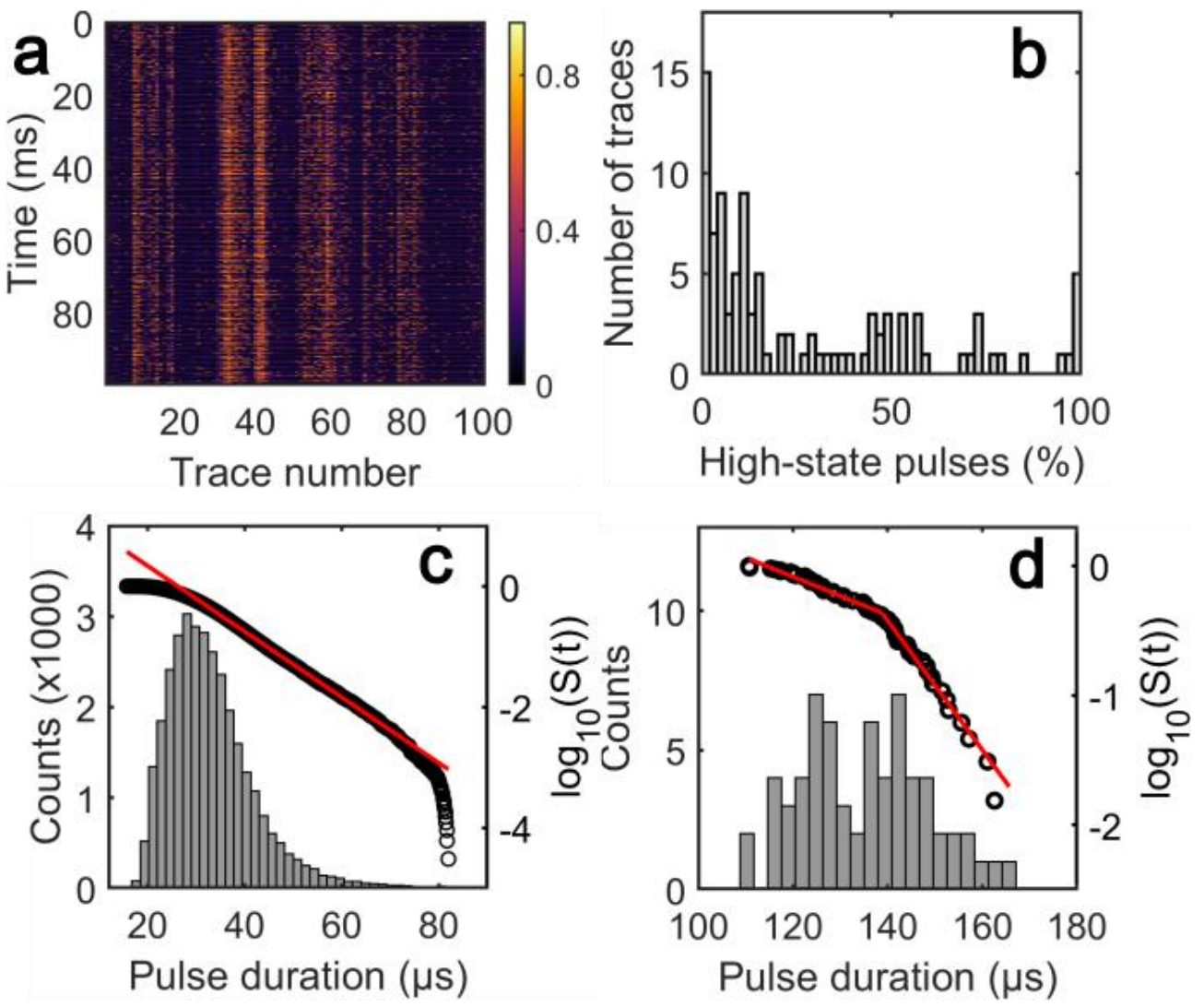

**Figure 4. Pulse excitation, and temporal memory dynamics.** Temporal map of 100 optical traces with duration of 100 ms (e). Probability distribution of high-state pulse occurrences across recorded traces (f). Pulse duration distributions of high-state pulses below one period (g) and between one and two periods (h): calculated (black circles, right axis) and fitted (red lines, right axis) survival functions.

3c. This behavior is attributed to the excitation of the high state, which occurs when the modulating current crosses all or most of the upward transition currents distribution, $I_{UP}$.

In Fig. 3d, we overlay the input electrical pulse with the low and high intensity optical responses. Notably, high-intensity overshoots arising from the normal low-intensity response are characterized by a much shorter rise time and persist for much longer than the electrical pulse. Their duration varies from pulse to pulse.

## Statistics and pulse duration

The sudden jump to high level power is faster (0.7 μs) than the speed allowed by direct modulation of the LD and it is attributed to carrier-photons dynamics, whit subsequent relaxation oscillations, smeared out by the photodiode bandwidth. Once turned on, the high-level state lasts longer than the modulating pulse duration, *i.e.* the output power is high long after the current is back to $I_{DC}$.

We acquire a set of 100 traces each with 100 ms duration. Acquisition time delay between traces (about 300 ms) is set by the communication speed with the oscilloscope and the number of points ($10^6$) per trace. Acquired traces are not consecutive and are analyzed statistically. We first analyze the occurrence of high-state pulses over all the acquired traces and then their duration distributions.

In Fig. 4a, all acquired traces are shown. We observe bursts of high-state pulses which last for all the 100 ms duration of one trace, alternating with traces with only low-state pulses and traces with varying percentage of both. Considering the large total time of acquisition, in the range of seconds, these variations are attributed to slow fluctuations of the system, *i.e.* small thermal changes.

For each trace, we calculated the percentage of high-state pulses relative to the total pulse count. The resulting distribution across all traces is shown in Fig. 4b A right skewed distribution is observed with almost 50% of the traces having less than 20% of high-state pulses. Above 20%, the data points are scattered uniformly across the horizontal axis, with the frequency rarely exceeding 2 traces per bin, except for an accumulation of 5 traces with 100% of high-state pulses.

The duration of high-state pulses, which vary from pulse to pulse, is measured from all the acquired traces. The vast majority (99%) of the high-state pulse durations fall inside a single modulation period (100 μs) and a few high-state pulses exceed one period and are triggered by the excitation pulse into the next period. We first discriminate the pulse durations below one period and plot their distribution in Fig. 4c. We observe a minimum pulse duration of about 15 μs, three times larger than the excitation pulse width, the peak of the distribution is located at 30 μs and a long tail extends up to ∼ 80 μs. The mean duration is 33 μs with a standard deviation of 9 μs.

We calculate the survival function of the high-state pulse duration, $S(t) = P(\Delta T > t)$, defined as the probability $P$ that a pulse duration $\Delta T$ lasts longer than a given time $t$, see Fig. 4c. The semilogarithmic plot reveals three distinct dynamical regimes. For lifetimes below approximately 25 μs, the survival function remains flat at $S(t) = 1$, validating the existence of a duration window where downward transitions are completely inhibited. Between 25 μs and 75 μs $\log[S(t)]$ exhibits a clear linear decay, so in this range $S(t) = \exp[-(t-t_0)/\tau]$, where $t_0$ is the dead time (or minimum pulse duration) and $\tau$ is the characteristic lifetime of the high state. The good fit we obtain provides definitive proof of an exponential survival probability, confirming that the state collapse in this regime is a memoryless, stochastic process governed by a characteristic lifetime of $\tau = 8$ μs and $t_0 = 26.2$ μs. Finally, the sharp plunge near 80 μs, close to the period length of 87 μs, corresponds to the total absence of pulse with duration longer than 82.3 μs. Pulses that last longer than this duration are triggered into the next period.

In Fig. 4d, we plot the statistical distribution of durations for high-state pulses that prolong into the next 100 μs cycle. Notice that the number of events is far smaller. We observe a bimodal structure with a local minimum of durations at 133 μs and a minimum pulse duration of ~ 110 µs. The semilogarithmic plot of $S(t)$ shows two linear regions, from 110 µs to 140 µs and from 140 µs to 160 µs. The exponential fit of $S(t)$ in these intervals, gives $t_{01}$ = 114 µs and $\tau_1$ = 29.3 µs and $t_{02}$ = 131.5 µs and $\tau_2$ = 7.9 µs, respectively.

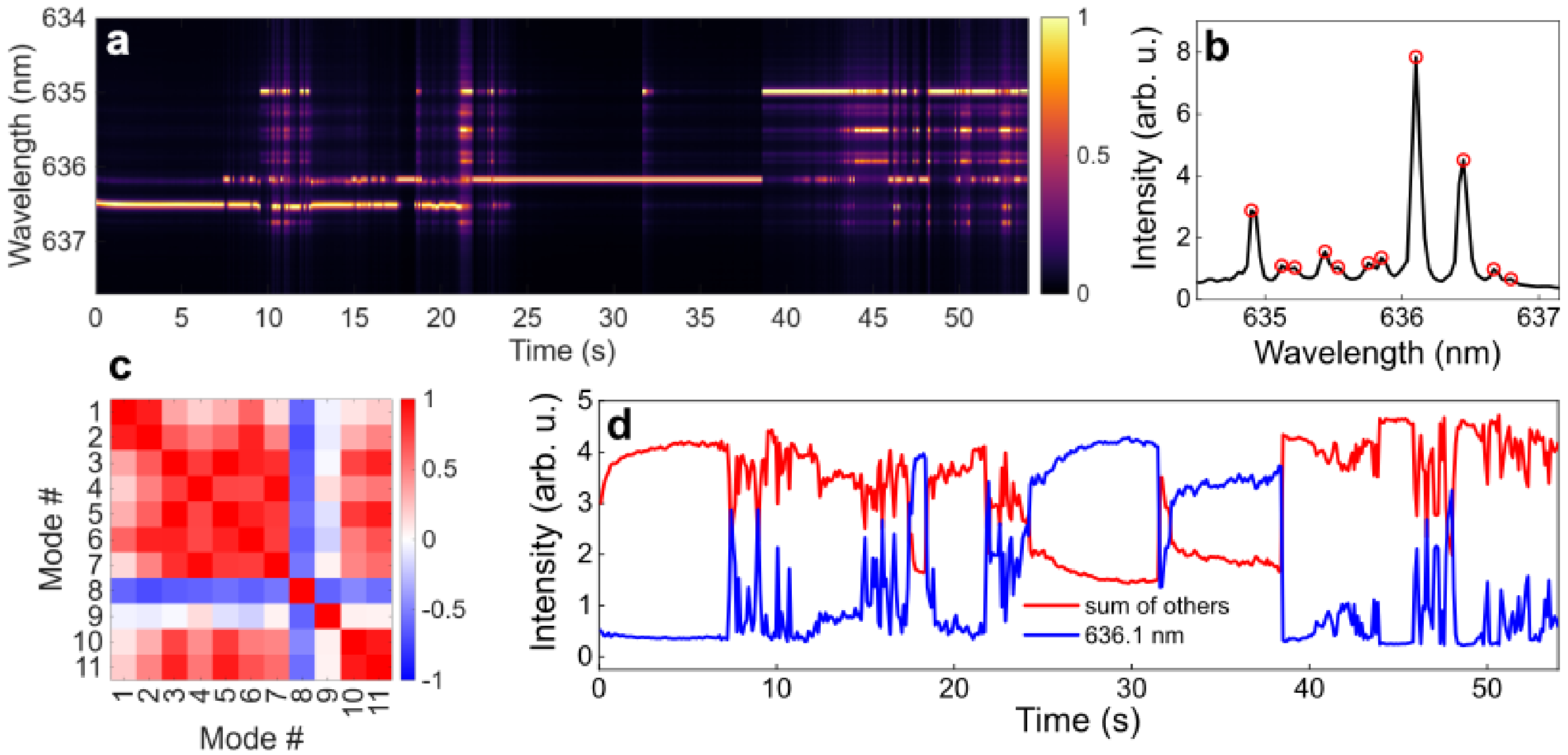

**Figure 4. Mode switching dynamics.** Emitted spectra acquired over a time window of 54 seconds at a fixed bias current of 245 mA (a). Sum of all detected spectra in a (black line) with detected peaks of emission (red circles), from mode 1 (634.56 nm) to mode 11 (636.45 nm) (b). Pearson correlation matrix of the temporal profiles of modes 1-11 (c). Temporal intensity traces of mode 8 (635.76 nm, blue line) and the sum of all other modes (red line) (d).

The arrival of the electrical pulse on a second cycle excites carriers, perturbs available optical gain and prompts the collapse of a first group of high-states pulses before reaching a duration of about ~ 133 µs. High-state pulses which survive this perturbation, enter the exponential, memoryless decay, resulting in a peak at 145 µs and a tail up to 160 µs.

**Mode Switching.**

For currents above the hysteresis range, we observed different spectral emission signatures even if the bias current was fixed. In order to investigating this behavior, the bias current was fixed at 245 mA, clear of the ($I_{DOWN}$, $I_{UP}$) interval, and emission spectra acquired at the fastest sampling rate allowed by our spectrometer, about every 110 ms (9 Hz).

In Fig. 5a, we present the time resolved emission plot, normalized to their maximum value, acquired over a time window of 54 seconds. The emission spectrum varies dynamically through sudden jumps between distinct operational configurations, transitioning from single dominant mode to broad emission comprising several simultaneous peaks in an erratic, random, fashion.

The total intensity, obtained by integration on the wavelength range, fluctuates randomly over the considered time interval, and the peak wavelength jumps among dominant modes, see Section 3 of Supplementary Information.

Summing up the spectra over the full duration yields the time-integrated spectral profile shown in Fig. 5b. We can see that rather than averaging out, emission shows eleven distinct longitudinal modes fairly resolved, dominated by Mode #1 (634.56 nm), Mode #8 (635.76 nm), and Mode #9 (636.1 nm). It is worth noticing here that none of the predominant modes derives from the pristine cavity LD spectrum whence for this driving current a dominant mode could be expected at 638 nm. The time resolved emission exhibits three key dynamical regimes: an initial phase (0–20 s) dominated by Mode #9 before entering rapid mode-hopping with Mode #8; a single-mode dominated plateau (24–38 s) where Mode #8 suppresses surrounding cavity emission; and a multi-mode relaxation phase (38–54 s) where Mode #8 collapses and power redistributes into the secondary spectrum with Mode #1 carrying more intensity than the rest.

To analyze the global interactions across the full observation duration without time-window discrimination, we compute the 11×11 Pearson correlation matrix across all mode pairs (Fig. 5c, indexed from Mode #1 to Mode #11). The matrix demonstrates that Mode 8 acts as the dominant competitor within the cavity, exhibiting widespread anti-correlation coefficients (between -0.5 and -0.7) across virtually all other longitudinal modes. Conversely, the secondary modes (Modes #1–7, #10, and #11) display positive mutual cross-correlations among one another, demonstrating that they co-exist and co-vary as a collective ensemble that fluctuates inversely to Mode #8. Mode #9 exhibits a near-zero overall correlation coefficient. This is understood at sight of its almost constant intensity outside the range when Mode #8 goes through its high-intensity period (24–43 s), and Mode #9 is all but switched off.

Finally, Fig. 5d directly illustrates this modal competition by plotting the temporal evolution of Mode #8 alongside the integrated intensity of all secondary modes. A clear anti-correlation is observed: whenever Mode #8 gains intensity, the secondary ensemble suffers simultaneous intensity loss. And this anticorrelation not only occurs with the ensemble but, as Fig. 4 c shows, it happens for every single mode: notice #8 column/row is the only negative one. This complementary exchange reflects cross-gain mode competition driven by coupled fluctuations in temperature, carrier density, and refractive index, which steer the laser across distinct emission states operating on fast (carrier) and slow (thermal) time scales.

**Discussion.**

The non-linear dynamics observed in our output-facet engineered RLD, encompassing optical hysteresis, stochastic excitability, spectral mode hopping, and anti-correlated modal behavior, demonstrate that disorder introduced directly onto the laser mirror transforms a standard LD into a self-contained, multi-stable dynamical system. Rather than requiring external optical feedback, optical injection, or multi-section cavities, these phenomena emerge directly from the internal interplay between complex scattered optical feedback and local carrier, thermal, and optical field dynamics.

To understand the microscopic origins of this non-linear behavior, we first consider the primary physical changes induced by the FIB milling process. The significant reduction in differential quantum efficiency compared to the pristine device, accompanied by an elevated threshold current, cannot be explained solely by increased mirror transmission losses. In fact, reduced mirror reflectivity alone would increase slope efficiency. Instead, the low quantum efficiency points directly to two main loss mechanisms: augmented internal scattering losses due to light backscattered into non-guided modes, and a marked reduction in internal quantum efficiency. The reduced internal quantum efficiency, coupled with the pronounced pre-threshold blue-shift, indicates both elevated carrier densities prior to threshold and the presence of non-radiative recombination centers at the milled facet, such as dangling bonds, lattice defect networks, and residual gold ion implantation.

These localized non-radiative recombination centers act as localized, pump-dependent thermal and carrier sinks. Injected current drives local carrier density and thermal dissipation, which directly governs the localized refractive index profile through both thermal dispersion and carrier-induced band filling or plasma effects. Because this localized refractive index profile governs light propagation within the laser cavity as well as the precise resonance and phase conditions of the backscattered waves from the randomized slot matrix, any change in local pump or optical power alters the complex feedback matrix. This reshapes the spatial optical field and spatial hole burning profile, which continuously feeds back into the local carrier density and heat dissipation.

As a consequence of this strongly coupled feedback loop, multiple self-consistent spatial configurations of carrier density, temperature, refractive index, and optical fields can satisfy the lasing threshold and boundary conditions at a single injection current. This multi-stability manifests experimentally across the distinct operational regimes observed.

In static current sweeps, the overlapping stability domains of two distinct states produce a hysteresis loop. The stochastic nature of the transition currents indicates that thermal noise and carrier fluctuations can kick the system across the effective energy barrier separating state basins of attraction.

Under square-pulse modulation, excitation into the high-power state triggers rapid carrier-photon dynamics, followed by state persistence long after the electrical pulse has ceased. The survival function of high-state lifetimes reveals a deterministic minimum residence time, followed by a memoryless exponential decay driven by stochastic carrier and phase noise triggering state collapse.

At higher constant bias current, cross-gain saturation drives dynamic competition among spatial states. We observed a primary mode acting as a dominant channel that strongly anti-correlates with the surrounding secondary modes. When local thermal or carrier fluctuations quench this primary mode, the refractive index profile instantaneously shifts, altering resonance conditions across the facet and allowing the collective ensemble of secondary modes to temporarily achieve threshold.

**Conclusions.**

In summary, our experimental evidence demonstrates that output facet FIB-milling can introduce controlled disordered feedback tightly coupled to local non-radiative recombination dynamics. The resulting carrier-thermal-optical feedback loop enables reproducible multistability and rich non-linear dynamics inside a single semiconductor laser die.

Specifically, we demonstrate counterclockwise optical hysteresis between distinct emission states, stochastic excitability under pulsed electrical modulation, spectral mode hopping, and anti-correlated modal dynamics at high bias currents. The hysteresis reveals bistability with stochastic switching between states, while the pulsed response exhibits high-power excursions that persist beyond the electrical excitation, with a well-defined minimum lifetime followed by an exponential decay. At higher bias currents, competition between optical modes leads to the intermittent suppression of a dominant mode and the emergence of secondary modes. Together, these observations establish that engineered disorder at the output facet can fundamentally reshape the dynamical behavior of a semiconductor laser without external optical feedback or injection.

This offers a compact, externally unperturbed platform for optical computing architectures, hardware-based information processing, and complex non-linear photonic signaling.

## Acknowledgements

This work has received funding from Ministerio de Ciencia, Innovación y Universidades through the project PSYNC (PID2024-158832NB-C21). We acknowledge support from the Spanish MICIU Severo Ochoa program for Centres of Excellence through grant CEX2024-001445-S/ funded by MICIU/AEI/ 10.13039/501100011033.

# Supplementary Information

## Section 1. Characterization of original Fabry-Perot laser diode.

The original devices are commercially available multimode laser diodes (Thorlabs HL6388MG) with emission wavelength at 635 nm. The device is delivered in a TO-56 can package, which is removed with a can opener (Thorlabs WR1). Figure S1a shows the opened device, revealing the circular TO-can housing, the electrical contacts, the heat sink, the substrate, and the laser chip, highlighted by the red rectangle. A magnified view of the laser chip is shown in Fig. S1b, where the anode contact is located on the top surface and the emitting area is highlighted in yellow. The dimensions of the laser chip are: 180 μm width, 100 μm height and 1.45 mm length.

The original LD is characterized under CW operation by measuring the output power and emission spectrum as a function of the bias current. The device is mounted on a temperature-controlled laser mount (Arroyo Instruments, 234B TEC TO-Can LaserMount), which is driven by an Arroyo Instruments 6305 ComboSource. The driver enables automated measurements via a PC running MATLAB.

The output light is collected using a microscope objective (20×, NA = 0.9), collimated with a lens (focal length of 2.54 cm), and then focused back either onto a calibrated photodiode (Thorlabs S130C), controlled via a Thorlabs PM100USB power meter, or into an optical fiber with a core diameter of 100 μm connected to a CCD spectrometer (Andor Shamrock 303) with a spectral resolution of 0.15 nm. All measurements are performed with the laser mount temperature maintained at 20 °C. Instrument control and measurement automation are implemented in MATLAB.

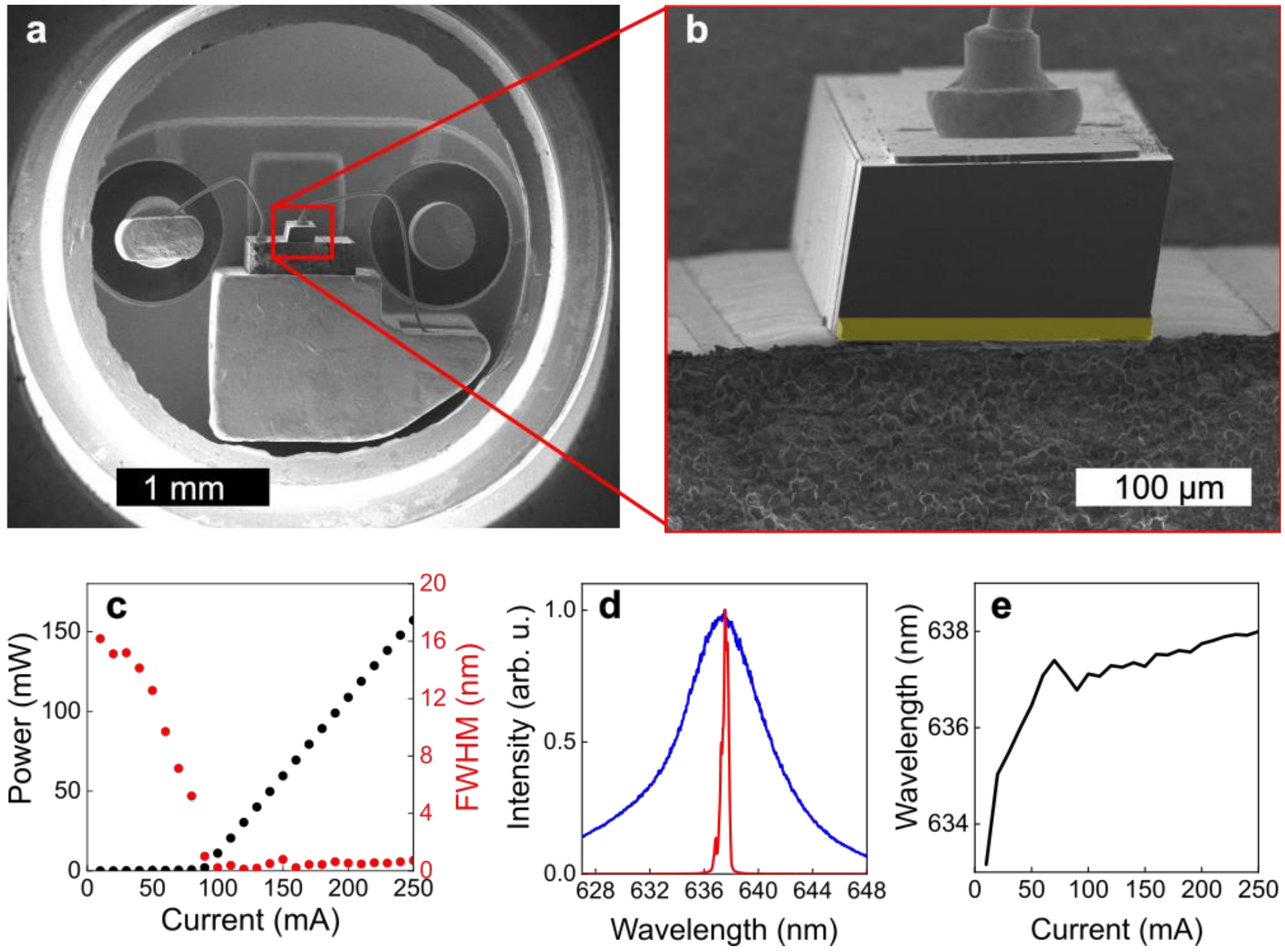

**Figure S1. Characterization of the original LD.** SEM images of the original LD, large view (a) and zoomed view (b). Output power (black dots, right axis) and FWHM linewidth (red dots, left axis) as a function of bias current (c). Emission spectra at 0.78 $I_{TH}$ (blue line) and at 2 $I_{TH}$ (red line) (d). Central wavelength of emission as a function of bias current (e).

These experimental set-ups are used for the CW characterization of the output power and spectra from the RLD.

In Fig. S1c, we show the output power and FWHM linewidth of the emission spectrum as a function of bias current. Power and spectral measurements are performed with the set-up described in the Methods section of the main text. From Fig. S1c, a threshold current $I_{\mathrm{TH}}$ = 90 mA and an efficiency slope $\eta_{\mathrm{SLOPE}}$ = 0.9 W/A are obtained. The differential quantum efficiency $\eta_{\mathrm{D}}$ is calculated as $\eta_{\mathrm{D}} = q \cdot \lambda \cdot \eta_{\mathrm{SLOPE}} / (h \cdot c)$ , where $q$ is the elementary charge, $\lambda$ is the central emission wavelength, $h$ is the Planck's constant and $c$ the speed of light in vacuum [ref]. This yields a differential quantum efficiency of $\eta_{\mathrm{D}}$ = 51 %. The FWHM linewidth decreases from approximately 16 nm below threshold to subnanometer values above threshold, ranging from 0.12 to 0.8 nm. Figure S1d shows representative emission spectra below and above threshold, measured at 0.8 $I_{\mathrm{TH}}$ and 2 $I_{\mathrm{TH}}$, respectively. Below threshold, the spectrum is broad, corresponding to amplified spontaneous emission. Above threshold, the spectrum narrows considerably, with an FWHM linewidth of 0.45 nm at 2·$I_{\mathrm{TH}}$. The spectrum exhibits a dominant peak accompanied by several weaker satellite peaks, consistent with the multimode emission of the device.

In Fig. S1e, we plot the central emission wavelength as a function of bias current. Below threshold, the wavelength increases from 633.1 nm to 637.4 nm, mainly due to Joule heating of the active region. Around the threshold current, a small blueshift of approximately 0.6 nm is observed. Above threshold, the wavelength increases approximately linearly with current, consistent with the thermal redshift typically observed in FP LDs.

## Section 2. Hysteresis characterization.

To assess the reproducibility of the hysteresis and quantify the cycle-to-cycle variability of the switching currents, we analyzed 100 consecutive current-sweep cycles. Figure S2a shows the distributions of the upward and downward switching currents, $I_{UP}$ and $I_{DOWN}$, respectively, obtained over 100 consecutive cycles. Their mean values are 200.6 mA and 192.6 mA, with standard deviations of 0.6 mA and 1.2 mA, respectively. No systematic drift is observed across successive cycles; instead, both transition currents fluctuate irregularly, with no consistent tendency to increase or decrease.

Figure S2b shows the distribution of the hysteresis width $W$, $W = I_{UP} - I_{DOWN}$, calculated for each individual cycle using the corresponding pair of switching currents. The distribution has a standard deviation of 1.16 mA and exhibits an asymmetric shape, reflecting the different spreads of the $I_{UP}$ and $I_{DOWN}$ distributions.

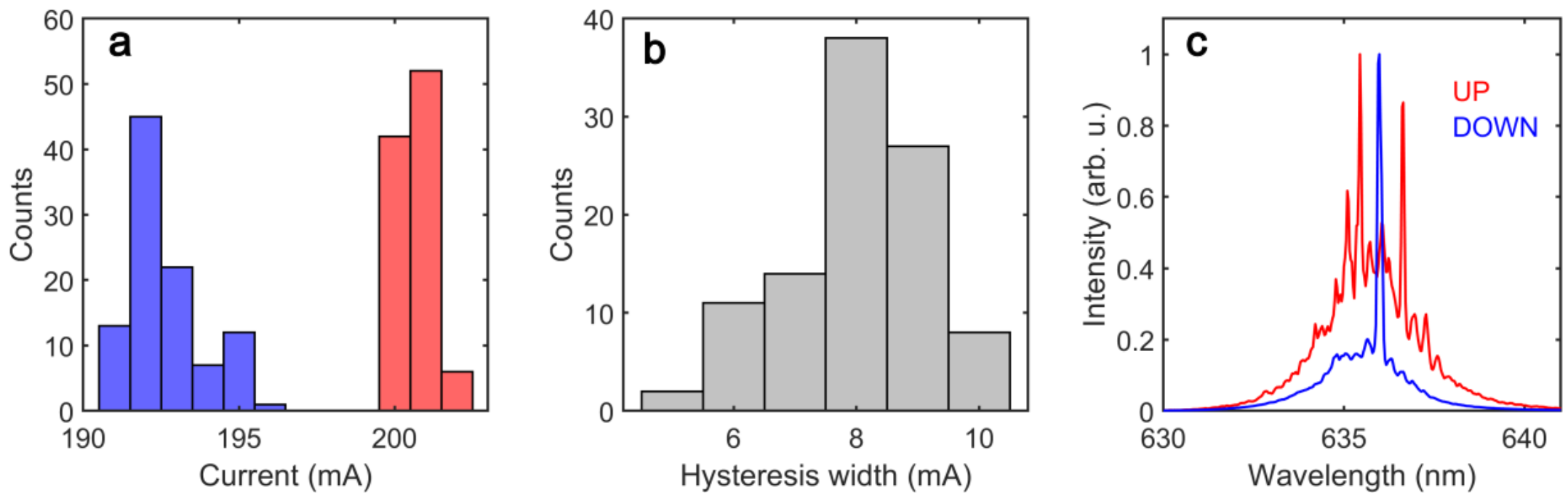

**Figure S2. Statistical characterization of hysteresis cycle.** Distributions of the upward ($I_{UP}$, red bars) and downward ($I_{DOWN}$, blue bars) switching currents, obtained over 100 consecutive current-sweep cycles (a). Distribution of the hysteresis width, calculated for each cycle (b). Normalized emission spectra measured at 200 mA during the upward (red) and downward (blue) current sweeps (c).

The spectral signatures associated with hysteresis are illustrated in Fig. S2(c), which compares the normalized emission spectra measured at 200 mA during the upward (red line) and downward (blue line) current sweeps. On the lower branch, during the upward sweep, the spectrum exhibits multimode emission, characterized by several narrow peaks superimposed on a broad spectral background. In contrast, on the upper branch, during the downward sweep, the spectrum is dominated by a single narrow peak, while the broad background remains present. This single-mode emission state, reached when the current exceeds $I_{UP}$, persists as the current decreases until $I_{DOWN}$ is crossed, whereupon multimode emission is restored.

Pulsed measurements described in the main text are obtained with the following modulation set-up. The laser mount and current and temperature controller are the same as those used for the CW measurements. The current driver supports modulation via an external voltage source at frequencies up to 325 kHz, with a nominal voltage-to-current conversion factor of 25 mA/V. An external pulse generator is connected to the current driver via a BNC cable.

The laser output is collected using a microscope objective (20×, NA = 0.9) and collimated with a lens with a focal length of 2.54 cm. The beam is then focused onto an amplified photodiode (Thorlabs PDA36A), which provides selectable gain and bandwidth settings. For these measurements, the gain is set to 0 dB and the bandwidth to 16 MHz. The photodiode output is recorded using a digital oscilloscope (Tektronix DPO3032) with a bandwidth of 500 MHz.

## Section 3. Mode switching.

The spontaneous spectral dynamics observed at a constant injection current of 245 mA are analyzed in this section. Figure S3a shows the spectra measured under this bias current; the data are the same as those presented in Fig. 5a, but are displayed as measured, without normalization to the maximum intensity. Figure S3b shows the total intensity, integrated over the wavelength range from 634 to 637 nm, as a function of time. The intensity trace is relatively smooth when a single spectral peak clearly dominates, as observed from 0 to 7 s and from 24 to 38.6 s. In contrast, a noisier profile is observed during intervals of continuous switching of the dominant mode, namely from 7 to 24 s and after 38.6 s. This behavior is consistent with changes in the system's dynamical state, with relatively stable emission and periods of intense switching persisting for several seconds. Transient perturbations can also occur during otherwise stable emission. For example, at 31.5 s, the dominant mode abruptly switches from Mode #8 (635.76 nm) to Mode #1 (636.1 nm) and then returns to Mode #8. At 38.6 s, Mode #1 becomes dominant, while competing with other modes, including Mode #8, until the end of the recorded trace.

The alternation between periods of modal competition and relatively stable emission is also evident in the evolution of the dominant peak wavelength shown in Fig. S3c. Stable emission corresponds to intervals in which the wavelength remains nearly constant, whereas periods of modal competition are characterized by sudden and repeated jumps between wavelengths. A closer inspection of Fig. S3c reveals that Mode #9 is associated with two closely spaced peak wavelengths, at 636.077 nm and 636.14 nm. These two wavelengths are distinguishable in the time-resolved data but are not resolved as separate features in the time-integrated spectrum in Fig. 5b, where their contributions merge into a single spectral feature.

The frequency-domain characteristics of the intensity trace in Fig. S3b, obtained using the fast Fourier transform (FFT), are shown in Fig. S3d. The FFT amplitude is largest at low frequencies, particularly between 0 and 1 Hz, and generally decreases towards higher frequencies. This indicates that slow intensity variations make a substantial contribution to the observed dynamics, consistent with the extended intervals of spectral persistence in Fig. S3a–c. Fast switching dynamics occurring at different rates contribute to the fluctuations observed at higher frequencies in the FFT spectrum. No single pronounced spectral peak is observed, suggesting that the intensity dynamics contain contributions across a range of frequencies rather than being dominated by a single frequency component.

In Fig. S3e, we show the histogram of the peak wavelengths whose temporal evolution is presented in Fig. S3c. Of the 11 peaks identified in Fig. 5b of the main text, five appear in the histogram, indicating that they become dominant at some point during the measurement. Three modes occur most frequently: Mode #1 (634.56 nm), Mode #8 (635.76 nm), and Mode #9 (636.1 nm). The remaining two modes, Mode #4 (635.1 nm) and Mode #7 (635.5 nm), appear much less frequently, with 23 counts and only one count, respectively. The two closely spaced wavelengths associated with Mode #9, discussed above in the context of Fig. S3c, fall within the same histogram bin and are therefore counted together.

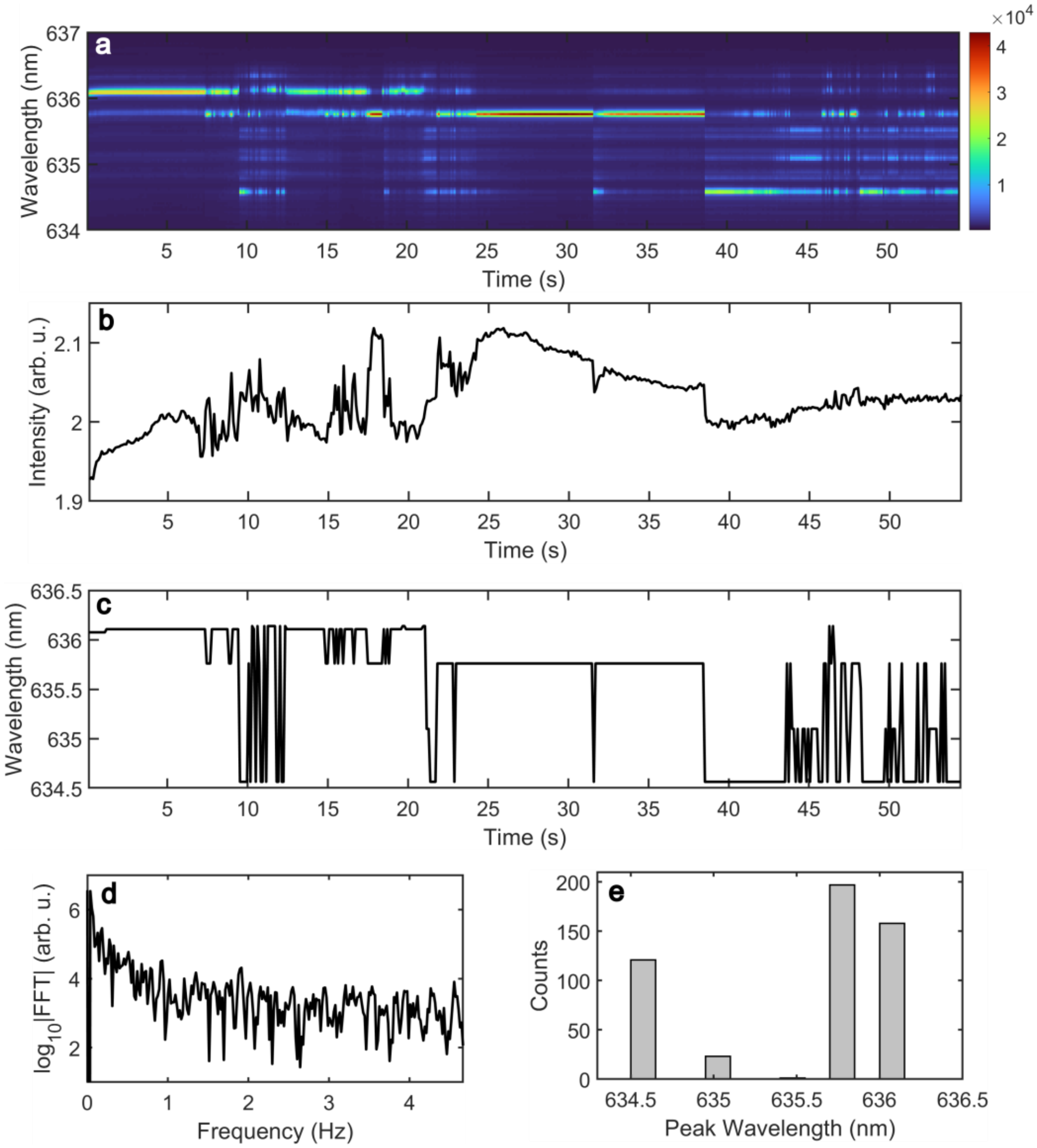

**Figure S3. Spectral dynamics at constant injection current of 245 mA.** Time-resolved emission spectrum (a). Total emission intensity integrated over the wavelength range as a function of time (b). Temporal evolution of the maximum peak wavelength (c). Fourier transform, in logarithmic scale, of the temporal trace of total emission intensity, plotted in panel b. Histogram of the dominant peak wavelengths recorded over the measurement interval (e).